\documentclass[11pt,a4paper,twoside]{JHEP3}

\usepackage{cite}
\usepackage{graphicx}
\usepackage{epsfig}
\usepackage{dcolumn}
\usepackage{bm}

\def \d {{\rm d}}

\newcommand{\be}{\begin{equation}}
\newcommand{\ee}{\end{equation}}

\newcommand{\beqn}{\begin{eqnarray}}
\newcommand{\eeqn}{\end{eqnarray}}

\newcommand{\pa}{\partial}
\newcommand{\pp}{{\it pp\,}-}
\newcommand{\ba}{\begin{array}}
\newcommand{\ea}{\end{array}}

\def \d {{\rm d}}

\def \bk {\mbox{\boldmath{$k$}}}

\title{Higher dimensional spacetimes with a geodesic, shearfree, twistfree and expanding null congruence\footnote{Proceedings of the XVII SIGRAV Conference, Turin, September 4--7, 2006.}}

\author{Marcello Ortaggio \\ 
        Dipartimento di Fisica, Universit\`a degli Studi di Trento \\ 
        and INFN, Gruppo Collegato di Trento \\
        Via Sommarive 14, 38050 Povo (Trento), Italy \\
        {E-mail: \email{ortaggio AT science.unitn.it}}}

\abstract{We present the complete family of higher dimensional spacetimes that admit a geodesic, shearfree, twistfree and expanding null congruence, thus extending the well-known $D=4$ class of Robinson-Trautman solutions. Einstein's equations are solved for empty space with an arbitrary
 cosmological constant and for aligned pure radiation. Main differences with respect to the $D=4$ case (such as the absence of type III/N solutions, related to ``violations'' of the Goldberg-Sachs theorem in $D>4$) are pointed out, also in connection with other recent works. A formal analogy with electromagnetic fields is briefly discussed in an appendix, where we demonstrate that multiple principal null directions of null Maxwell fields
 are necessarily geodesic, and that in $D>4$ they are also shearing if expanding.}

\dedicated{Dedicated to Massimo Pauri on the occasion of his nomination as {\em Professor Emeritus}.}

\begin{document}

\section{Introduction}
\label{intro}

Thanks to the correspondence between geometrical properties of
null geodesics and optical properties of the gravitational field,
the study of ray optics
\cite{Sachs61,EK,NP,piraniintro,penrosebook2} has played a major
role in the construction and classification of exact solutions of
Einstein's equations \cite{Stephanibook}. This applies in
particular to solutions representing gravitational radiation. Remarkably, the fundamental families of ``spherical'' and
``plane-fronted'' gravitational waves found, respectively, by
Robinson and Trautman \cite{RobTra60,RobTra62} and by Kundt
\cite{Kundt61,Kundt62} are invariantly defined in terms of
geometric optics. Relying on formal analogies with the theory of
electromagnetic radiation, Robinson-Trautman solutions are
characterized by the existence of a geodetic, non-twisting,
non-shearing and expanding null congruence, whereas Kundt metrics
(which include \pp waves) require a non-twisting and non-shearing
congruence but with vanishing expansion. In both cases, the
shear-free condition implies that the corresponding spacetimes are
algebraically special (at least in vacuum or in the presence of
``sufficiently aligned'' matter), due to the Goldberg-Sachs
theorem \cite{GolSac62,NP}. In fact, explicit solutions of all
special Petrov types are known within both classes of geometries
\cite{RobTra60,RobTra62,Kundt61,Kundt62,Stephanibook}.

The geometric optics approach was naturally developed in $D=4$
General Relativity. On the other hand, in recent years gravity in
more than four spacetime dimensions has become an active area of
ongoing interdisciplinary studies. It is thus now interesting to
investigate possible extensions of
the above concepts to arbitrary dimensions $D>4$. Certain properties of higher dimensional Kundt
solutions have been analyzed in
\cite{Coleyetal03,Obukhov04,ColHerPel06,Coleyetal06} (these in
particular contain all spacetimes with vanishing scalar curvature
invariants \cite{Coleyetal03,Coleyetal04vsi,Coleyetal06}). It is
the purpose of our contribution to summarize the systematical
derivation of the $D>4$ Robinson-Trautman class of solutions
performed in \cite{PodOrt06} and to discuss the main features of such
spacetimes and recent related developments. We shall focus here on the
case of vacuum spacetimes, with a possible cosmological constant,
and of aligned pure radiation. All our results will be local.

Aspects of geometric optics in higher dimensions have been also
studied in \cite{FroSto03,Pravdaetal04,LewPaw05}, and
``violations'' of the standard $D=4$ Goldberg-Sachs theorem in
$D>4$ have been pointed out. For example, the principal null
directions of $D=5$ rotating black holes (which are of type D) are
geodesic but shearing \cite{FroSto03,Pravdaetal04}. Furthermore, in $D>4$
vacuum spacetimes of type III or N, a multiple principal null
direction with expansion necessarily has also non-zero shear
(still being geodesic) \cite{Pravdaetal04}. This implies in particular that there do not
exist shearfree, twistfree and expanding (i.e., Robinson-Trautman)
vacuum solutions of type III or N when $D>4$, as opposed to the
$D=4$ case \cite{RobTra60,RobTra62,Stephanibook}. Our results,
obtained with a method different from that of \cite{Pravdaetal04}, will be in agreement with this
conclusion. Furthermore, we shall point out that also
electromagnetic ``null'' fields share similar properties in $D>4$.

The paper is organized as follows. In section~\ref{sec-geom} we
present the general form of the Robinson-Trautman line element
under purely geometrical requirements on the optical scalars. In
section~\ref{Einstein} we derive the explicit solution to
Einstein's equations within such a setting. In
section~\ref{sec-discussion} we discuss general
features of the obtained spacetimes, such as the algebraic
structure of the corresponding Weyl tensor, whereas we concentrate on the
special case of vacuum solutions in section~\ref{sec-vacuum}. We
present concluding remarks in section~\ref{sec-conclusions}.
Appendix~\ref{app-optics} summarizes geometric optics definitions
in higher dimensions, mainly following \cite{Pravdaetal04} (cf.
also \cite{FroSto03,LewPaw05}). Appendix~\ref{app-Maxwell} studies
optical properties of higher dimensional null Maxwell fields and
shows that their rays are necessarily geodesic, and that they must
be shearing if expanding.

\section{Geometrical assumptions: spacetimes with a geodesic, shearfree and twistfree
null congruence} \label{sec-geom}

Let us consider a generic $D$-dimensional spacetime ($D\ge4$). A
null congruence with tangent vector field $\tilde k^\alpha$ is
(locally) orthogonal to a family of null hypersurfaces
$u(x)=\mbox{const}$ (i.e., $\tilde k_\alpha=-f(x)u_{,\alpha}$,
with $g^{\alpha\beta}u_{,\alpha}u_{,\beta}=0$) if and only if
$\tilde k_{[\alpha;\beta}\tilde k_{\rho]}=0$.\footnote{The
implication $\tilde k_\alpha=-f(x)u_{,\alpha}\Rightarrow\tilde
k_{[\alpha;\beta}\tilde k_{\rho]}=0$ is obvious. The
converse follows from the Frobenius theorem, see
e.g.~\cite{Stephanibook}.} The latter condition, in turn, is
equivalent to $\tilde k^\alpha$ being geodesic and twistfree, in
the sense that the twist matrix of~\cite{Pravdaetal04} vanishes
(see appendix~\ref{app-optics}). Given such a congruence, the
rescaled tangent field $k_\alpha=-u_{,\alpha}$ will also be null,
twistfree and geodesic and, in addition, affinely parameterized
(i.e., $k_{\alpha;\beta}k^\beta=0$). Now, it is natural to take
the function $u$ itself (constant along each ray) as one of the
coordinates, so that $k_\alpha=-\delta^u_\alpha$ and $g^{uu}=0$.
As for the remaining coordinates, we use the affine parameter $r$
along the geodesics generated by $k^\alpha$, and ``transverse''
spatial coordinates $(x^2, x^3, \ldots , x^{D-1})$ which are
constant along these null geodesics
(cf.~\cite{Brinkmann25,RobTra62}). This further implies
$k^\alpha=\delta^\alpha_r$, that is $g^{ur}=-1$ and
$g^{ui}=0$. Therefore, the covariant line element can be
written as
\be
 \d s^2=g_{ij}\left(\d x^i+ g^{ri}\d u\right)\left(\d x^j+ g^{rj}\d u\right)-2\d u\d r-g^{rr}\d u^2 ,
 \label{geo_metric}
\ee
where the metric coefficients are, for now, arbitrary functions of all the coordinates $(x,u,r)$ (hereafter, $x$ stands for all the transverse coordinates $x^i$ and lowercase latin indices range as $i=2,\ldots,D-1$).
Useful relations between the covariant and contravariant metric coefficients, to be employed in the sequel, are
\be
 g^{ri}= g^{ij}g_{uj} , \qquad g^{rr}=-g_{uu}+g^{ij}g_{ui}g_{uj} , \qquad g_{ui}= g^{rj}g_{ij} ,
 \label{cov_contra}
\ee
while $g_{rr}=0=g_{ri}$, $g_{ur}=-1$.
In this coordinate system, it is also easy to see that
\be
 k_{\alpha;\beta}=\frac{1}{2}g_{\alpha\beta,r} .
\ee
For later purposes, it is convenient (cf.~\cite{RobTra62,PodOrt06}) to define an auxiliary $(D-2)$-dimensional spatial metric $\gamma_{ij}$ by
\be
 \gamma_{ij}=p^2g_{ij} , \qquad p^{2(2-D)}\equiv\det g_{ij}=-\det g_{\alpha\beta} ,
 \label{gamma and P}
\ee so that $\det\gamma_{ij}=1$. Then one can express the
generalized optical scalars expansion and shear
\cite{FroSto03,Pravdaetal04} (cf. appendix~\ref{app-optics})
associated to $k^\alpha$ simply as \cite{PodOrt06}\footnote{The definitions of the scalars $\mbox{Tr}(\sigma^2)$ and $\theta$ in~(\ref{scalars}) hold only when an {\em affine} parameter is used along~$k^\alpha$.} 
\beqn
 & & \mbox{Tr}(\sigma^2)\equiv k_{(\alpha;\beta)}k^{\alpha;\beta}-\frac{1}{D-2}(k^\alpha_{\;;\alpha})^2=\frac{1}{4}\gamma^{li}\gamma^{kj}\gamma_{ki,r}\gamma_{lj,r} , \nonumber \label{scalars} \\
 & & \theta\equiv\frac{1}{D-2}k^\alpha_{\;;\alpha}=-(\ln p)_{,r} .
\eeqn

Now, imposing the condition that the congruence $k^\alpha$ is {\em shear-free}, $\mbox{Tr}(\sigma^2)=0$, eq.~(\ref{scalars}) leads to
\be
 \gamma_{ij,r}=0 ,
\ee
since there always exists a frame in which $\gamma^{ij}$  is diagonal, with strictly positive eigenvalues.

In summary, the line element of any spacetime admitting a
hypersurface orthogonal (i.e., geodesic and twistfree) shear-free
null congruence $\mbox{\boldmath$k$}=\pa_r$ can be written in
the form (\ref{geo_metric}), with $g_{ij}=p^{-2}\gamma_{ij}$; the
matrix $\gamma_{ij}$ is unimodular and independent of $r$, while
$p$, $g^{ri}$ and $g^{rr}$ are arbitrary functions of $(x,u,r)$.
Note that such a metric is left invariant by the following
coordinate transformations (which do not change the family of null
hypersuperfaces $u=\mbox{const}$ nor the affine character of the
parameter~$r$)
\be
 x^i=x^i(\tilde x,\tilde u) , \qquad u=u(\tilde u), \qquad r=r_0(\tilde x,\tilde u)+\tilde r/\dot u(\tilde u) .
 \label{freedom}
\ee

\section{Integration of Einstein's equations}
\label{Einstein}

After deriving the above line element~(\ref{geo_metric}) (with
eq.~(\ref{gamma and P})), one has to integrate Einstein's
equations  $R_{\alpha\beta}-\frac{1}{2}Rg_{\alpha\beta}+\Lambda
g_{\alpha\beta}=8\pi T_{\alpha\beta}$ to determine the unknown
metric coefficients $g_{ij}$, $g^{ri}$ and $g^{rr}$. Here we 
concentrate on the case of vacuum spacetimes ($T_{\alpha\beta}=0$) and
of aligned pure radiation ($T_{\alpha\beta}=\Phi^2 k_\alpha
k_\beta$), while the cosmological constant $\Lambda$ is arbitrary.
In the coordinate system introduced above this means that only the
$T_{uu}=\Phi^2$ component can be non-vanishing. Then, recalling
the line element~(\ref{geo_metric}) and the
relations~(\ref{cov_contra}), Einstein's equations effectively
reduce to the following set of equations: $R_{rr}=0$, $R_{ri}=0$,
$R_{ij}=\frac{2}{D-2}\Lambda g_{ij}$,
$R_{ur}=-\frac{2}{D-2}\Lambda$, $R_{ui}=\frac{2}{D-2}\Lambda
g_{ui}$ and $R_{uu}=\frac{2}{D-2}\Lambda g_{uu}+8\pi \Phi^2$.

Since we have assumed that twist and shear are identically vanishing, the generalized Sachs equation governing the rate of change of the expansion \cite{LewPaw05,PodOrt06} (namely $(D-2)\theta_{,\alpha} k^\alpha+\mbox{Tr}(\sigma^2)+(D-2)\theta^2+\mbox{Tr}(A^2)=-R_{\alpha\beta}k^\alpha k^\beta$) implies that for the $R_{rr}$ component we have simply $R_{rr}=-(D-2)(\theta_{,r}+\theta^2)$. Hence the integration of $R_{rr}=0$ singles out two alternative solutions. The first arises when $\theta=0$ which, by~(\ref{scalars}), is equivalent to $p=p(x,u)$. This possibility corresponds to the Kundt class of non-expanding spacetimes, considered in \cite{Coleyetal03,Obukhov04,ColHerPel06,Coleyetal06} for $D>4$. But we are here interested in the alternative case $\theta\not=0$ of an expanding vector field $k^\alpha$, for which one finds $\theta^{-1}=r+r_0(x,u)$. The arbitrary function $r_0$ can always be set to zero by a transformation~(\ref{freedom}), so that
\be
 \theta=\frac{1}{r} .
\ee From eq.~(\ref{scalars}) we can thus factorize
$p=r^{-1}P(x,u)$, where $P$ is an arbitrary function. It is also
convenient to rescale the transverse metric by introducing
$h_{ij}=P^{-2}\gamma_{ij}$, so that the relation  (\ref{gamma and
P}) becomes
\be
 g_{ij}=p^{-2}\gamma_{ij}=r^2P^{-2}\gamma_{ij}=r^2h_{ij}(x,u) .
 \label{hspatial}
\ee
This specifies the $r$-dependence of $g_{ij}$ in the metric (\ref{geo_metric}). Note that $P^2=(\det h_{ij})^{1/(2-D)}$.

One should now proceed to integrate all the remaining Einstein
equations. Since this is lengthy, we just summarize here the main
results and refer to \cite{PodOrt06} for technical details. First,
imposing (a subset of) the Einstein equations and using an appropriate coordinate
transformation one can always set
\be
 g^{ri}=0 .
\ee
Then, one finds that at any given $u=u_0=\mbox{const}$ each
$(D-2)$-dimensional spatial metric $h_{ij}(x,u_0)$ must be an
Einstein space (in $D=5$ this implies that $h_{ij}$ is a 3-space
of constant curvature); also, the $u$-dependence of $h_{ij}$ must
be factorized out in a conformal factor. Namely, \beqn
 & & {\mathcal R}_{ij}=\frac{{\mathcal R}}{D-2}h_{ij} , \label{constrijr} \\
 & & h_{ij}=P^{-2}(x,u)\gamma_{ij}(x) , \qquad \mbox{ where } \quad \det \gamma_{ij}=1 , \label{hijs}
\eeqn in which ${\mathcal R_{ij}}$ is the Ricci tensor associated
with the metric $h_{ij}$, and ${\mathcal R}$ the corresponding
Ricci scalar. As a well-known fact, any two-dimensional metric
$h_{ij}$ satisfies eq.~(\ref{constrijr}), so that this is
identically satisfied in the special case $D=4$. Therefore for
$D=4$ eq.~(\ref{constrijr}) puts no restriction on $h_{ij}$ (in
particular,  the scalar curvature ${\mathcal R}$ of $h_{ij}$ can
generally depend both on $u$ and on the spatial coordinates~$x$).
On the other hand, for any $D>4$, eq.~(\ref{constrijr}) tells us
(via the contracted Bianchi identities) that ${\mathcal R}$ can
not depend on the $x$ coordinates, i.e.
\be
 {\mathcal R}={\mathcal R}(u) \qquad (D>4) .
\ee As a consequence, the next equation to be solved (which
controls the $u$-dependence of $P$) differs crucially in the $D=4$
and $D>4$ cases, namely one has \beqn
 & & \frac{1}{2}({\mathcal R}_{,i}h^{ij})_{,j}-({\mathcal R}_{,i}h^{ij})(\ln P)_{,j}+6\mu(\ln P)_{,u}-2\mu_{,u} =16\pi n^2 \qquad (D=4) , \label{RTD4} \\
 & & (D-1)\mu(\ln P)_{,u}-\mu_{,u} =\frac{16\pi n^2}{D-2}  \qquad (D>4),  \label{constrspec2}
\eeqn where $n=n(x,u)$ and $\mu=\mu(u)$ are arbitrary functions.
The former characterizes the pure radiation term, which must take
the form
\be
 T_{uu}=\Phi^2=r^{2-D}n^2(x,u) ,
\label{matter} \ee whereas the latter enters the last metric
coefficient $g^{rr}=-g_{uu}\equiv 2H$, given by
\be
2H=\frac{{\cal R}}{(D-2)(D-3)}-2r(\ln P)_{,u}-\frac{2\Lambda}{(D-2)(D-1)}\,r^2-\frac{\mu(u)}{r^{D-3}} . \label{Hfin}
\ee
Note that the first two terms in
eq.~(\ref{RTD4}) represent (one-half of) the covariant Laplace
operator on a 2-space with metric $h_{ij}$ (applied to ${\mathcal
R}$). Renaming $\mu=2m$ we thus obtain $\frac{1}{2}\Delta{\mathcal
R}+12m(\ln P)_{,u}-4m_{,u}=16\pi n^2$, which is the standard form
of the Robinson-Trautman equation \cite{Stephanibook}. We will not
discuss the case $D=4$ any longer here since it is already
well-known.

\section{Properties of the solutions in $D>4$}

\label{sec-discussion}

To summarize, the Robinson-Trautman class of solutions in $D>4$
with aligned pure radiation, as obtained above, reads
\be
 \d s^2=r^2P^{-2}\gamma_{ij}\d x^i\d x^j-2\d u\d r-2H\d u^2 ,
 \label{geo_metric fin}
\ee with eqs.~(\ref{hspatial}), (\ref{constrijr}), (\ref{hijs}),
(\ref{constrspec2}), (\ref{matter}) and (\ref{Hfin}). Notice that
using the coordinate freedom remaining from (\ref{freedom}),
namely the reparametrization of $u$ ($\dot u>0$),
\be
u=u(\tilde u), \quad r=\tilde r/\dot u(\tilde u), \quad \hbox{so
that} \quad \tilde P=P\dot u, \ \tilde{\cal R}={\cal R}{\dot u}^2,
\ \tilde \mu=\mu{\dot u}^{D-1}, \ \tilde n^2=n^2\dot u^D,
 \label{freedomspec}
\ee ($\tilde{\cal R}$ is indeed the Ricci scalar of the rescaled
metric $\tilde h_{ij}={\dot u}^{-2}h_{ij}$) we may achieve further
useful simplification of the metric (\ref{geo_metric fin}),
(\ref{Hfin}). For example, we can always put $\tilde\mu$ or (in
$D>4$) $\tilde{\cal R}$ to be a constant. Also, transformations of
the coordinates $x^i=x^i(\tilde x)$ can be used to change the form
of the spatial metric $h_{ij}$. In any case, for fixed $r$ and $u$
the metric $\tilde h_{ij}$ can be any Riemannian Einstein space
(see eq.~(\ref{constrijr})), the $u$-dependence of this family
being governed solely by $P$. The variety of possible metrics
$h_{ij}$ is thus huge. For example, in addition to the simplest
case when $h_{ij}$ is a space of constant curvature, if ${\cal
R}>0$ and $5\le D-2\le 9$ one can take any of the infinite number
of non-trivial compact Einstein spaces presented in \cite{Bohm98}.

One can also compute the Weyl tensor corresponding to the
metric~(\ref{geo_metric fin}). Its non-zero components read
\cite{PodOrt06} \beqn
 & & C_{ruru}=-\mu(u)\frac{(D-2)(D-3)}{2r^{D-1}} , \qquad C_{riuj}=\mu(u)\frac{(D-3)}{2r^{D-3}}h_{ij} , \nonumber \\
 & & C_{ijkl}=r^2{\cal R}_{ijkl}-2r^2\left( \frac{{\cal R}(u)}{(D-2)(D-3)}-\frac{\mu(u)}{r^{D-3}}\right)h_{i[k}h_{l]j} ,\label{Weyl}\\
 & & C_{uiuj}=2HC_{riuj} ,   \nonumber
\eeqn
where ${\cal R}_{ijkl}$ is the Riemann tensor associated to $h_{ij}$.
Using a suitable frame based on the null vectors
\be
 \mbox{\boldmath$k$}=\pa_r , \qquad \mbox{\boldmath$l$}=-\pa_u+H\pa_r
 \label{null_vec}
\ee (such that $\mbox{\boldmath$k$}\cdot\mbox{\boldmath$l$}=1$),
it is straightforward to see that the above coordinate components
give raise only to frame components of boost weight zero. The Weyl
tensor is thus of type~D (unless vanishing) in the classification
of \cite{Coleyetal04,Milsonetal05}. This should be contrasted with
the case $D=4$, in which all algebraically special types are
allowed. Type O (conformally flat) spacetimes are possible only if
$\mu=0$, which implies a vanishing pure radiation field
(cf.~eq.~(\ref{constrspec2})) and therefore that only constant
curvature spacetimes can occur in this case.

\section{Vacuum solutions}

\label{sec-vacuum}

The special case of vacuum Robinson-Trautman spacetimes is given
by $T_{uu}=0$, i.e. $n=0$ in eq.~(\ref{constrspec2}). In this
case, one has to consider separately the two cases $\mu\neq 0$ and
$\mu=0$.

\subsection{Case $\mu\neq 0$}

\label{subsec-vac1}

When $\mu\neq 0$, it can always be set to a constant by a
rescaling~(\ref{freedomspec}). Eq.~(\ref{constrspec2}) thus
reduces to ${\mu P_{,u}=0}$. Consequently, in this case the
function $P$ must be independent of~$u$, and thus must therefore
be also $h_{ij}$. It follows, in particular, that ${\cal R}$ is a
constant. Unless now ${\cal R}=0$, one can choose the
transformation~(\ref{freedomspec}) to normalize ${\cal R}=\pm
(D-2)(D-3)$. The corresponding Robinson-Trautman geometries are
thus fully characterized by the line element~(\ref{geo_metric
fin}) with the simple function
\be
 2H=K-\frac{2\Lambda}{(D-2)(D-1)}\,r^2-\frac{\mu}{r^{D-3}}  \qquad (K=0, \pm 1) ,
\label{Hvacuum} \ee and they clearly admit $\pa_u$ as a Killing
vector. As mentioned above, the spatial metric
$h_{ij}=P^{-2}(x)\gamma_{ij}(x)$ can describe {\em any} Einstein
space with scalar curvature ${\cal R}=K(D-2)(D-3)$. When this
space is compact, such a family of solutions describes various
well-known static black holes in Eddington-Finkelstein
coordinates. In particular, if the horizon has constant curvature
one obtains Schwarzschild-Kottler-Tangherlini black holes
\cite{Tangherlini63}, for which the line element can always be
cast in the form \cite{PodOrt06}
\be
 \d s^2=r^2\left(1+\textstyle{\frac{1}{4}}K\delta_{kl}x^kx^l\right)^{-2}\delta_{ij}\d x^i\d x^j-2\d u\d r-2H\d u^2 .
 \label{geo_metric spec2}
\ee In addition, there are generalized black holes
\cite{Birmingham99,GibIdaShi02prl,GibHar02} with various horizon
geometries (e.g., those presented in \cite{Bohm98} for $K=+1$),
non-standard asymptotics and, possibly, non-spherical horizon
topology. As we have seen in section~\ref{sec-discussion}, all
these solutions are of type D, and the components~(\ref{Weyl})
of the Weyl tensor become \beqn
 & & C_{ruru}=-\mu\frac{(D-2)(D-3)}{2r^{D-1}} , \qquad C_{riuj}=\mu\frac{(D-3)}{2r^{D-3}}h_{ij} , \nonumber \\
 & & C_{ijkl}=r^2{\cal R}_{ijkl}-2r^2\left( K-\frac{\mu}{r^{D-3}}\right)h_{i[k}h_{l]j} ,\label{Weylvac1}\\
 & & C_{uiuj}=2HC_{riuj} .    \nonumber
\eeqn

\subsection{Case $\mu=0$}

\label{subsec-vac2}

In the exceptional case ${\mu=0}$, eq.~(\ref{constrspec2}) is
identically satisfied in vacuum, and one can not conclude that $P$
is independent of $u$. One can still rescale ${\cal R}(u)$ to be a
constant ${\cal R}=K(D-2)(D-3)$, ${K=0, \pm 1}$, so that in this
case the line element~(\ref{geo_metric fin}) contains the
characteristic function
\be
 2H=K-2r(\ln P)_{,u}-\frac{2\Lambda}{(D-2)(D-1)}\,r^2 .
\label{Hvacuum2}
\ee
Now the only non-vanishing components of the type D Weyl tensor are
\be
 C_{ijkl}=r^2{\cal C}_{ijkl} ,
\label{Weylvac2} 
\ee 
where ${\cal C}_{ijkl}$ is the Weyl tensor of $h_{ij}$.
These spacetimes degenerate to type O (thus
to constant curvature since vacuum) when $C_{ijkl}=0$, which is
equivalent to having a transverse spatial metric $h_{ij}$ of
constant curvature~$K$. In particular, this is necessarily the
case in $D=5$ (cf. section~\ref{Einstein}).

Now, recall from eq.~(\ref{hijs}) that
$h_{ij}(x,u)=P^{-2}(x,u)\gamma_{ij}(x)$, i.e. the possible
$u$-dependence of the line element~(\ref{geo_metric fin}) is
contained only in $P$. In the simplest case of a factorized
function $P(x,u)=P_1(x)P_2(u)$, this $u$-dependence is obviously
removable with a transformation~(\ref{freedomspec}). The
corresponding solutions are therefore equivalent to the metrics of
subsection~\ref{subsec-vac1} with $\mu=0$. In general, when the $u$-dependence is not factorized, the
solutions~(\ref{Hvacuum2}) are instead presented in a somewhat
implicit form, in the sense that one has still to specify an
Einstein metric $h_{ij}$ such that it depends on $u$ only via an
overall conformal factor. Finding an explicit spacetime of the
type~(\ref{Hvacuum2}) with a non-trivial $u$-dependence (in $D>5$,
since the trivial case $D=5$ has been already discussed above)
could be in principle all but straightforward. Notice, however,
the following: if we think of $h_{ij}$ as a family of
$(D-2)$-dimensional Riemannian Einstein metrics parametrized by
$u$, it is obvious that all such Einstein spaces are conformally
related. We can thus take any known Einstein space and obtain from
it a conformally related Einstein space with an appropriate
$u$-dependence in the conformal factor. In fact, conformal mapping
of Einstein spaces on Einstein spaces were studied thoroughly by
Brinkmann \cite{Brinkmann24,Brinkmann25} (see also, e.g.,
\cite{Eisenhart49,petrov}). In particular, four-dimensional
Riemannian Einstein spaces which admit a conformal
(non-homothetic) map on Einstein spaces must be of constant
curvature \cite{Brinkmann24,Brinkmann25}. Therefore by eq.~(\ref{Weylvac2}) also in $D=6$
(which means that $h_{ij}$ is four-dimensional)
solutions~(\ref{Hvacuum2}) trivially reduce to constant curvature
spacetimes (Minkowski or (anti-)de~Sitter) if $P(x,u)$ is non-factorized. On the other hand, in $D\ge 7$ we can employ
the ``canonical'' form
\cite{Brinkmann24,Brinkmann25,Eisenhart49,petrov} of Einstein
spaces which can be mapped conformally on other Einstein spaces to
construct an explicit, non-trivial solution of the
form~(\ref{Hvacuum2}). For example, in $D=7$ one of the simplest metrics we
can think of is the following (we take
$(z,\tau,\rho,\theta,\phi)$ as our $(x^2,\ldots,x^6)$ coordinates)
\beqn
 & & K=-1 , \qquad P(u,z,\rho,\theta)=f(u,z)^{-1/2}(\rho^2\sin\theta)^{-1/5} , \\
 & & h_{ij}\d x^i\d x^j=f(u,z)\Bigg[\d z^2+\left(1-\frac{2m}{\rho}-\frac{\rho^2}{l^2}\right)\d\tau^2 \nonumber \\
 & & \hspace{3cm} {}+\left(1-\frac{2m}{\rho}-\frac{\rho^2}{l^2}\right)^{-1}d\rho^2+\rho^2(\d\theta^2+\sin^2\theta\d\phi^2)\Bigg], \label{einstein_h} \\
 & & f(u,z)=\frac{4b(u)e^{2z/l}}{l^2[e^{2z/l}-b(u)]^2} \label{einstein_f},
\eeqn where $m$ and $l$ are constants and $b(u)>0$ is an arbitrary
function.\footnote{We clearly used the four-dimensional Euclidean
Schwarzschild-de~Sitter solution as a ``seed'' metric for the
five-dimensional Einstein space~(\ref{einstein_h}),
(\ref{einstein_f}).} This $u$-dependence turns out to have a
specific geometrical meaning \cite{PraPraOrt07} which
invariantly characterizes the spacetime and more general
solutions~(\ref{Hvacuum2}) (as opposed to
solutions~(\ref{Hvacuum})). The associated Weyl
tensor is non-zero and one can show, e.g., that
$C_{\mu\nu\rho\sigma}C^{\mu\nu\rho\sigma}=48m^2f^{-2}r^{-4}\rho^{-6}$.
Note also that one can set $\Lambda=0$, if desired.

\section{Conclusions}
\label{sec-conclusions}

We have presented the complete family of higher dimensional
spacetimes that contain a hypersurface orthogonal, non-shearing
and expanding congruence of null geodesics, and that satisfy
Einstein's equations with an arbitrary cosmological constant and
aligned pure radiation. In particular, we have discussed vacuum
solutions. There appear fundamental differences with respect to
the standard ${D=4}$ family of Robinson-Trautman solutions and,
remarkably, $D>4$ Robinson-Trautman spacetimes can be only of type
D or O. This is in agreement with the previous results of
Ref.~\cite{Pravdaetal04}, which proved (via a study of the Bianchi
identities) that the multiple principal null congruence of $D>4$
type N and type III vacuum spacetimes must have non-zero shear
when it has non-zero expansion, so that type N and type III can
not occur in the vacuum Robinson-Trautman class. For the case
of non-twisting null congruences, we have generalized this
conclusion to include a non-vanishing cosmological constant, and
to prove that in fact any algebraic type different from D and O is
forbidden for vacuum spacetimes with an expanding but non-shearing
null congruence.

It is interesting to remark that $D=4$ Robinson-Trautman solutions
were originally constructed so as to model spherical gravitational
radiation \cite{RobTra60,RobTra62}. They were thus required to
share certain algebraic and optical properties with
corresponding solutions of Maxwell's equations describing
electromagnetic waves, in particular ``null'' fields. In contrast,
it turned out that null (type N) gravitational fields are not
possible for $D>4$ vacuum Robinson-Trautman solutions. We show in
appendix~\ref{app-Maxwell} that expanding null electromagnetic fields
provide a natural counterpart to this in the higher dimensional
Maxwell theory.

\appendix

\section{Optical scalars in higher dimensions}

\label{app-optics}

In a $D$-dimensional spacetime, one can introduce a frame of
vectors
$(\mbox{\boldmath$k$},\mbox{\boldmath$l$},\mbox{\boldmath$m$}^{(i)})$
(with $i=2,\ldots,D-1$) which satisfy the  ``orthonormality''
relations \be \mbox{\boldmath$k$}\cdot\mbox{\boldmath$l$}=1 ,
\qquad
\mbox{\boldmath$m$}^{(i)}\cdot\mbox{\boldmath$m$}^{(j)}=\delta_{ij}
, \qquad
\mbox{\boldmath$k$}\cdot\mbox{\boldmath$k$}=\mbox{\boldmath$l$}\cdot\mbox{\boldmath$l$}=\mbox{\boldmath$k$}\cdot\mbox{\boldmath$m$}^{(i)}=\mbox{\boldmath$l$}\cdot\mbox{\boldmath$m$}^{(i)}=0
, \ee so that $\mbox{\boldmath$k$}$ and $\mbox{\boldmath$l$}$ are
null and the $\mbox{\boldmath$m$}^{(i)}$ are spacelike.
On such a frame, the metric takes the simple form
$g_{\mu\nu}=2k_{(\mu}l_{\nu)}+\delta_{ij}m^{(i)}_\mu m^{(j)}_\nu$.
Also, one can decompose the covariant derivative of the null
vector $k_\mu$ as \cite{Pravdaetal04}
\be
  k_{\mu;\nu}= K_{11}k_{\mu}k_\nu+K_{10}k_\mu l_\nu+K_{1i}k_\mu m^{(i)}_{\nu}
  +K_{i1} m^{(i)}_{\mu}k_{\nu}+K_{i0} m^{(i)}_{\mu}l_{\nu}+K_{ij}m^{(i)}_{\mu}m^{(j)}_{\nu} ,
  \label{cov_der}
\ee
where, obviously,
\beqn
  & &  K_{11}\equiv k_{\mu;\nu}l^\mu l^\nu , \qquad K_{10}\equiv k_{\mu;\nu}l^\mu k^\nu ,
  \qquad K_{1i}\equiv k_{\mu;\nu}l^{\mu}m^{(i)\nu} , \nonumber \\
  & & K_{i1}\equiv  k_{\mu;\nu}m^{(i)\mu}l^{\nu}, \qquad K_{i0}\equiv k_{\mu;\nu}m^{(i)\mu}k^\nu , \qquad K_{ij}\equiv k_{\mu;\nu}m^{(i)\mu}m^{(j)\nu} .
\eeqn

We are particularly interested in the case of a {\em geodesic}
vector field $k^\mu$. It follows from eq.~(\ref{cov_der}) that
$k_{\mu;\nu}k^\nu=K_{10}k_\mu+K_{i0} m^{(i)}_{\mu}$, so that
$k^\mu$ is geodesic if and only if $K_{i0}=0$. In addition,
$k^\mu$ is affinely parametrized if also $K_{10}=0$. When $k^\mu$
is geodesic the spatial matrix $K_{ij}$ acquires a special meaning
since it is then invariant under null rotations preserving $k^\mu$
(and it simply rescales with boost weight one under a boost in the
$\mbox{\boldmath$k$}$-$\mbox{\boldmath$l$}$ plane). It can thus be
used to invariantly characterize geometric properties of the null
congruence of integral curves of $k^\mu$. To this effect, let us
decompose $K_{ij}$ into its tracefree symmetric part, its trace
and its antisymmetric part as
\be
 K_{ij}=\sigma_{ij}+\theta\delta_{ij}+A_{ij} ,
\ee 
where 
\be
 \sigma_{ij}\equiv K_{(ij)}-\frac{\mbox{Tr}\,K}{D-2}\delta_{ij}=\sigma_{ji} ,
 \qquad \theta\equiv\frac{\mbox{Tr}\,K}{D-2} , \qquad A_{ij}\equiv
 K_{[ij]}=-A_{ji} . \label{opt_matrices}
\ee We shall refer to $\sigma_{ij}$ and $A_{ij}$ as the {\em
shear} and {\em twist} matrix, respectively, and to $\theta$ as
the {\em expansion} scalar. Along with $\theta$, one can now
construct other scalar quantities out of $k_{\mu;\nu}$ which are
invariant under null and spatial rotations with fixed $k^\mu$,
e.g. the shear and twist scalars given by $\mbox{Tr}(\sigma^2)$
and $\mbox{Tr}(A^2)$. If in addition
$k^\mu$ is affinely parametrized, i.e. $K_{10}=0$, one can express
the thus defined optical scalars solely in terms of $k^\mu$ as
\beqn
 & & \mbox{Tr}(\sigma^2)=k_{(\mu;\nu)}k^{\mu;\nu}-\frac{1}{D-2}\left(k^\mu_{\;;\mu}\right)^2 , \\
 & & \theta=\frac{1}{D-2}k^\mu_{\;;\mu} , \\
 & & \mbox{Tr}(A^2)=-k_{[\mu;\nu]}k^{\mu;\nu} .
\eeqn 
Notice, in particular, that
$\sigma_{ij}=0\Leftrightarrow\mbox{Tr}(\sigma^2)=0$ and
$A_{ij}=0\Leftrightarrow\mbox{Tr}(A^2)=0$.

Using the decomposition~(\ref{cov_der}) and
eq.~(\ref{opt_matrices}) it is also easy to see that
\be
 k_{[\mu;\nu}k_{\rho]}=K_{i0}m^{(i)}_{[\mu}l_\nu k_{\rho]}+A_{ij}m^{(i)}_{[\mu}m^{(j)}_{\nu}k_{\rho]}
 ,
\ee hence $k_{[\mu;\nu}k_{\rho]}=0$ if and only if $k^\mu$ is
geodesic ($K_{i0}=0$) and twistfree ($A_{ij}=0$).

All the above is a natural extension of the geometric optics
formalism developed and widely employed in $D=4$ General
Relativity
\cite{Sachs61,RobTra62,Kundt61,Kundt62,EK,NP,piraniintro,Stephanibook,penrosebook2}.

\section{Rays of null Maxwell fields in higher dimensions}

\label{app-Maxwell}

Given a ``null'' frame
$(\mbox{\boldmath$k$},\mbox{\boldmath$l$},\mbox{\boldmath$m$}^{(i)})$
as in appendix~\ref{app-optics}, the Maxwell 2-form $F_{\mu\nu}$ representing any electromagnetic field
can be decomposed in terms of its frame components (of different
boost weight) as $F_{\mu\nu}=2F_{0i}l_{[\mu}m^{(i)}_{\nu]}+2F_{01}l_{[\mu}k_{\nu]}+F_{ij}m^{(i)}_{[\mu}m^{(j)}_{\nu]}+2F_{1i}k_{[\mu}
m^{(i)}_{\nu]}$ \cite{Coleyetal04vsi,Milsonetal05,Milson04}. We shall focus here on Maxwell fields that, in an
appropriate frame, have only negative boost weight components,
i.e. $F_{0i}=F_{01}=F_{ij}=0$. Under this assumption the 2-form
$F_{\mu\nu}$ takes the simple form
\be
 F_{\mu\nu}=2F_{1i}k_{[\mu}m^{(i)}_{\nu]} , \label{null_Maxwell} 
\ee 
and (by analogy with the four-dimensional case
\cite{syngespec,Mariot54,robinsonnull,Stephanibook,penrosebook2}) it can then be
referred to as a type N or {\em null} Maxwell field, $k^\mu$ being
its multiple principal null direction. In fact, in any $D\ge 4$ dimension a
Maxwell field has vanishing (zeroth-order) scalar invariants if
and only if it is of type N \cite{Coleyetal04vsi}.

Let us now consider consequences of the vacuum Maxwell equations
$F^{\mu\nu}_{\ \ \ ;\nu}=0$ and $F_{[\mu\nu;\rho]}=0$ on optical
properties of the aligned null direction $k^\mu$. Using the
decomposition~(\ref{cov_der}), from the contractions
$F^{\mu\nu}_{\ \ \ ;\nu}k_\mu=0$ and $ F_{[\mu\nu;\rho]}k^\mu
m^{(i)\nu}m^{(j)\rho}=0$ one finds, respectively,
\be
 F_{1i}K_{i0}=0 , \qquad F_{1i}K_{j0}=F_{1j}K_{i0} .
 \label{Maxwell1}
\ee 
Contracting the second of eqs.~(\ref{Maxwell1}) with $K_{i0}$
and using the first of eqs.~(\ref{Maxwell1}) it follows
\be
 K_{i0}=0 ,
\ee i.e. $\bk$ is {\em geodetic}. Then, combining $ F^{\mu\nu}_{\
\ \ ;\nu}m^{(i)}_\mu=0$ and $F_{[\mu\nu;\rho]}l^\mu
m^{(i)\nu}k^\rho=0$ (and using
$k^\mu_{\;;\mu}=K_{10}+(D-2)\theta$) one gets
\be
 2F_{1i}\sigma_{ij}-(D-4)\theta F_{1j}=0 .
\ee 
This equation has now somewhat different consequences in $D=4$
and $D>4$. For $D=4$, one necessarily has $\sigma_{ij}=0$ (this
is most easily seen in a frame such that $\sigma_{ij}=\mbox{diag}(\sigma_{22},-\sigma_{22})$), i.e. the
aligned null direction $k^\mu$ must be shearfree
\cite{Mariot54,robinsonnull} (irrespective of the value of $\theta$). In
contrast, for $D>4$ if $k^\mu$ is expanding ($\theta\neq 0$) then
it must be also shearing ($\sigma_{ij}\neq 0$).

It is worth remarking that aligned null directions of type N Weyl
tensors display very similar properties \cite{Pravdaetal04}.



\providecommand{\href}[2]{#2}\begingroup\raggedright\endgroup

\end{document}